\documentclass[review]{elsarticle}
\usepackage{lineno}
\usepackage{hyperref}
\usepackage{amsmath}
\usepackage{amsfonts}
\usepackage{amssymb}
\usepackage{mathrsfs}
\usepackage{graphicx}
\usepackage{float}
\usepackage{color}
\usepackage{empheq}

\modulolinenumbers[5]

\allowdisplaybreaks

\journal{Wave Motion}

\begin{document}

\begin{frontmatter}

\title{The Malyuzhinets---Popov diffraction problem revisited}

\author[mymainaddress,mysecondaryaddress]{Ekaterina A. Zlobina\corref{mycorrespondingauthor}}
\cortext[mycorrespondingauthor]{Corresponding author}
\ead{ezlobina2@yandex.ru}

\author[mymainaddress,mysecondaryaddress,mythirdlyaddress]{Aleksei P. Kiselev}
\ead{kiselev@pdmi.ras.ru}

\address[mymainaddress]{St. Petersburg Department of V. A. Steklov Institute of Mathematics of the Russian Academy of Sciences, Fontanka Emb. 27, St. Petersburg 191023, Russia}
\address[mysecondaryaddress]{St. Petersburg State University, Universitetskaya Emb. 7-9, St. Petersburg, 199034, Russia}
\address[mythirdlyaddress]{Institute for Problems of Mechanical Engineering of Russian Academy of Sciences, Vasilievsky Ostrov Bolshoy Prospect 61, St. Petersburg 199178, Russia}

\begin{abstract}
In this paper, the high-frequency diffraction of a plane wave incident along a planar boundary turning into a smooth convex contour, so that the curvature undergoes a jump, is asymptotically analysed.
An approach modifying the Fock parabolic-equation method is developed.
Asymptotic formulas for the wavefield in the illuminated area, shadow, and the penumbra are derived.
The penumbral field is characterized by novel and previously unseen special functions that resemble Fock's integrals.
\end{abstract}


\end{frontmatter}


\section{Introduction}

\begin{figure}
\noindent\centering{
\includegraphics[width=.6\textwidth]{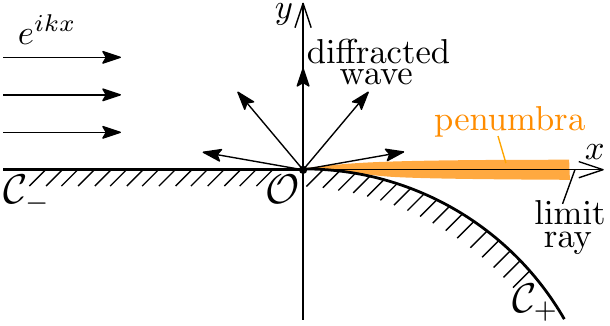}
}
\caption{Geometry of the problem}
\label{f1}
\end{figure}

In a remarkable paper \cite{PopovLiPar_en}, Alexey V. Popov addressed the high-frequency diffraction of a plane wave incident along a planar boundary (with the Neumann condition), passing into a parabola at its apex (Fig \ref{f1}).
He aimed at describing a diffracted cylindrical wave that emerges from the point of non-smoothness $\mathcal{O}$ of the contour $\mathcal{C}$ in accordance with the Geometrical Theory of Diffraction (GTD) \cite{Keller62,BorovikovKinber_en,James} and, by means of virtuoso calculations, obtained an explicit formula for it.
Popov's research was inspired by Malyuzhinets' paper \cite{Malyu}, in which this problem was first clearly formulated and qualitatively investigated.
The Malyuzhinets---Popov problem has much in common with the so-called Fock problem (see \cite{Fock0,FockDiffProbl_en,Brown,BabichKirp_en}), which consists of investigation of a high-frequency field in the vicinity of a tangency point in the diffraction of a plane wave by a smooth, convex obstacle.
A significant qualitative difference is the presence of a diffracted cylindrical wave in the illuminated region instead of a reflected one.
Fock's approach (inspired by a physical idea due to Leontovich) is based on introduction of an approximate equation called the \textit{parabolic equation}.
Fock dealt with a simplified problem that admits an explicit but rather complicated solution that he explored asymptotically.
He was attracted to the description of the field at a short distance from the obstacle, primarily in the vicinity of the limit ray, for which he developed a spectacular analytical technique.

A. Popov ingeniously used a complex combination of the parabolic-equation method, Malyuzhinets' technique and Kirchoff's approach to derive an expression for the diffracted wave.
However, he did not address the wavefield in the shadow and the penumbra surrounding the limit ray that separates illuminated and shadowed areas (Fig. \ref{f1}).
We observed a certain inaccuracy in the paper by A. Popov \cite{PopovLiPar_en}, which apparently did not affect the expression for the diffracted wave, but would impact upon a detailed description of the field in the penumbra.

There are two reasons for revisiting this problem.
First, for half a century, steady interest has been seen in the description of the effects of high-frequency diffraction by contours in which the curvature is neither strictly positive nor strictly negative (see, e.g., \cite{PopovLiPar_en,BabSmyshl86_en,BabSmyshl87_en,HewOckSmyshlWM19,Kazakov03,Nakamura89,OckTew12,OckTew21,PopovMM79_en,PopovMM82_en,PopovMM86_en,Smyshl91_en,SmyshlKamAA22_en,KirpPhil_Acoustics_en}).
A. Popov's work is historically the first in which such phenomena were quantitatively approached and one of the few (see, e.g., \cite{KirpPhil_Acoustics_en}) where the sign of curvature changes not in a smooth manner, but in a jump.

Second, the effects of a jump in the curvature (as well as that of weaker singularities; see \cite{ZloKisBigH_en}) do not allow a description from simple model problems because they do not exist.
These effects were studied for non-tangential incidence by heuristic approaches, such as the Kirchhoff method (see, e.g., \cite{James,Weston62,Senior72,KiselevRogoff,ZloKisRadEl22_en}), and by the boundary-layer theory \cite{KamKell,ZloKisWM20}.
For problems with tangential incidence, no close inspection based on the boundary-layer method is available.

We address the Malyuzhinets---Popov problem as a relatively simple problem with a tangential incidence on the boundary with a jump in the curvature.
We use a systematic boundary-layer technique, which goes back to the research of Fock (see \cite{Fock0}, \cite[ch. 7]{FockDiffProbl_en}) and was further developed by Brown \cite{Brown} and Babich and Kirpichnikova \cite[ch. 6]{BabichKirp_en}.
Similar to these studies, we examined wavefields in a small neighborhood of the singular point of the boundary.
We had to overcome significant analytical difficulties in deriving an expression for the diffracted wave (which agrees with the findings of A. Popov \cite{PopovLiPar_en}) and in investigating the Fresnel field and an analog of Fock's background field in the penumbra.
In addition, we found an expression for the field in deep shadow that matches with the creeping waves.
At all stages, we carefully estimated the remainder terms, which allowed us to reliably indicate the domains of validity of the resulting expressions.

\section{Formulation of the problem}

We consider the wavefield $u$ governed above the contour $\mathcal{C}$  by the Helmholtz equation
\begin{equation}\label{Helm}
	\left(\partial^2_x + \partial^2_y + k^2 \right)u = 0
\end{equation}
with large wavenumber $k$,
\begin{equation}
 k \gg 1,
\end{equation}
and satisfying the Neumann boundary condition
\begin{equation}\label{cond}
	\left. \partial_n u \right|_\mathcal{C} = 0.
\end{equation}
Here, $\partial_n$ is the derivative along the inner normal to $\mathcal{C}$.

The contour consists of the flat part $\mathcal{C}_-$ touching the smooth curved part $\mathcal{C}_+$ at point $\mathcal{O}$ (Fig. \ref{f1}).
Near $\mathcal{O}$ the curvature $\mbox{\ae}$ of contour $\mathcal{C}$ has the following form:
\begin{equation}\label{curv}
	\mbox{\ae}(x) = h\theta(x),
\end{equation}
where
\begin{equation}
	\theta(x) = \begin{cases}
		1, x > 0,\\
		0, x \le 0,
	\end{cases}
\end{equation}
is the Heaviside step function, and $h \neq 0$ is the magnitude of the jump of curvature.

The total wavefield $u$ is the sum of the incident plane wave $u^\text{inc} = e^{ikx}$ that travels left to right along $\mathcal{C}_-$ towards $\mathcal{O}$ and the outgoing wave $u^\text{out}$:
\begin{equation}\label{sum}
	u = u^\text{inc} + u^\text{out}.
\end{equation}
The problem is to describe the outgoing wave.

The problem under consideration is similar to that of Fock in that the limit ray is surrounded by several boundary layers, where the wavefield behavior differs.
The principal distinction is that the diffracted wave $u^\text{dif}$ emerges at the non-smoothness point $\mathcal{O}$, for which the GTD \cite{BorovikovKinber_en,James} predicts the following approximation far above the limit ray:
\begin{equation}\label{cylindrical}
    u^\text{dif} \approx A(\varphi; k) \frac{e^{ikr}}{\sqrt{kr}}, \quad kr \gg 1.
\end{equation}
Here, $A(\varphi; k)$ is a diffraction coefficient and $(r,\varphi)$ denotes the classical polar coordinates centred at $\mathcal{O}$:
\begin{equation}
	x = r\cos\varphi, \quad y = r\sin\varphi, \quad 0 \le r < \infty, \, -\pi \le \varphi < \pi.
\end{equation}
Unlike A. Popov, who was interested exclusively in the diffracted wave, we provide asymptotic descriptions of wavefields in boundary layers separating the illuminated region from the deep shadow, as shown in Fig. \ref{TotalPic}.
These descriptions employ functions resembling Fock's integrals \cite{FockDiffProbl_en,BabichKirp_en}.

\section{Parabolic-equation approach}
\subsection{Coordinates $(s,n)$}
\begin{figure}
	\noindent\centering{
	\includegraphics[width=.5\textwidth]{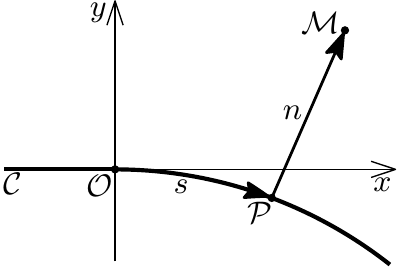}}
	\caption{Coordinates $s$ and $n$}
	\label{f2}
\end{figure}

It is natural to characterize the position of an observation point $\mathcal{M}$ near point $\mathcal{O}$ by coordinates $(s,n)$, where  $n \ge 0$ is the length of the perpendicular from $\mathcal{M}$ to the contour $\mathcal{C}$ and $s$ is the length of the arc between $\mathcal{O}$ and the foot of the perpendicular, as shown in Fig. \ref{f2}.
These coordinates are orthogonal; however, in contrast to the classical case \cite{FockDiffProbl_en,BabichKirp_en}, the mapping $(x,y) \mapsto (s,n)$ is not smooth.

Furthermore, $s$ and $n$ and, accordingly, $x$ and $y$ will be small in comparison with the curvature radius $h^{-1}$.
We use the following elementary relation which results from \eqref{curv} by a simple calculation:
\begin{subequations}\label{x,y->s,n}
	\begin{empheq}[left={\empheqlbrace}]{align}
		&x = s + \theta(s) \left(hns - \frac{h^2s^3}{6} + O\left(h^3s^3(n+hs^2) \right)\right),\\
		&y = n - \theta(s) \left(\frac{hs^2}{2} + O\left(h^2s^2(n+hs^2) \right) \right).
    \end{empheq}
\end{subequations}
We also use the expression for the Laplacian \cite{BabichKirp_en}:
\begin{equation}\label{preparab}
    \partial^2_x + \partial^2_y = \frac{1}{(1 + \varkappa(s)n)^2}\partial^2_s - \frac{n\varkappa'(s)}{(1 + \varkappa(s)n)^3}\partial_s + \partial^2_n + \frac{\varkappa(s)}{1 + \varkappa(s)n}\partial_n.
\end{equation}

\subsection{Reduction to the problem for parabolic equation}
We start with the observation (traceable to Leontovich and Fock) that, as in the case of the tangential incidence of a plane wave at a smooth obstacle, in the vicinity of point $\mathcal{O}$
\begin{equation}\label{2exp}
	u^\text{inc} = e^{ikx} = e^{iks} V,
\end{equation}
where $V$ oscillates slower than the exponentials.
This inspired us to follow the classical research strategy \cite{FockDiffProbl_en,Brown,BabichKirp_en} based on separating out a rapidly oscillating factor and introducing stretched coordinates, leading to a problem for the so-called parabolic equation.

We introduce dimensionless stretched coordinates by
\begin{equation}\label{stretch}
	\sigma = (h^2k/2)^{\frac{1}{3}}s, \quad \nu = (2hk^2)^{\frac{1}{3}} n,
\end{equation}
with the same powers of $k$ as in the original definition of Leontovich and Fock \cite{FockDiffProbl_en}.
The dimensionless large parameter
\begin{equation}
	k/h \gg 1
\end{equation}
of our problem is in agreement with that of Fock.
We seek the outgoing wavefield in a form similar to that of the Leontovich--Fock ansatz \cite{FockDiffProbl_en}:
\begin{equation}\label{U}
	u^\text{out} = e^{iks} U(\sigma, \nu) = e^{iks}\left(U_0(\sigma, \nu) + \ldots\right),
\end{equation}
where the \textit{attenuation factor} $U$ oscillates slower than $e^{iks}$ and the dots stand for smaller-order terms with respect to the large parameter $k/h$.

Formulas connecting coordinates $(\sigma, \nu)$ with the Cartesian coordinates follow from \eqref{x,y->s,n}:
\begin{subequations}\label{x,y->s,n_str}
	\begin{empheq}[left={\empheqlbrace}]{align}
		&k(x-s) = \theta(\sigma) \left(\nu\sigma - \frac{\sigma^3}{3} + O\left(\left(\frac{h}{k}\right)^{\frac{2}{3}}\sigma^3(\nu+\sigma^2) \right)\right), \label{x->s} \\
		&(2h)^{\frac{1}{3}} k^{\frac{2}{3}} y = \nu - \theta(\sigma) \left(\sigma^2 + O\left(\left(\frac{h}{k}\right)^{\frac{2}{3}}\sigma^2(\nu+\sigma^2) \right) \right).
	\end{empheq}
\end{subequations}
These formulas entail useful relations involving polar coordinates $r=\sqrt{x^2+y^2}$ and $\varphi = \arctan y/x$
\begin{gather}\label{r}
	k(r - s) = \frac{1}{2}\left(\frac{\nu^2}{2\sigma} +  \theta(\sigma)\left(\nu\sigma - \frac{\sigma^3}{6} \right)\right) + O\left[\left(\frac{h}{k}\right)^{\frac{2}{3}}\!\! \left(\sigma^3(\nu+\sigma^2) + \sigma\left(\nu-\frac{\sigma^2}{3}\right)^2 \right) \right],\\
	\varphi = \left(\frac{2h}{k}\right)^{\frac{1}{3}}\frac{\nu - \theta(\sigma)\sigma^2}{2\sigma} + O\left[\frac{h}{k} \left(\sigma(\nu+\sigma^2) + \frac{(\nu-\sigma^2)^2}{\sigma} \right)  \right].\label{phi}
\end{gather}

With the help of \eqref{x->s}, we rewrite the factor $V$ in \eqref{2exp} as
\begin{equation}\label{V}
	V(\sigma, \nu) = 1 - \theta(\sigma) + \theta(\sigma) e^{i\left(\sigma\nu - \frac{\sigma^3}{3}\right)} + O\left(\left(\frac{h}{k}\right)^{\frac{2}{3}}\sigma^3(\nu+\sigma^2) \right)
\end{equation}
and, following Fock, require the smallness of the remainder terms.
This implies the restrictions on distance from the singular point $\mathcal{O}$ which can be summarized as
\begin{equation}\label{upperlim}
    \sigma \ll (k/h)^{\frac{2}{15}}, \quad \nu \ll (k/h)^{\frac{4}{15}}.
\end{equation}
Further research refers to the area characterized by the conditions \eqref{upperlim}.\footnote{Limitations for the corresponding area in the Fock problem, presented in \cite{BabichKirp_en}, were different because the speed of propagation was assumed nonconstant.}

Rewriting \eqref{preparab} in stretched coordinates \eqref{stretch}, expanding the coefficients in powers of $(h/k)^{\frac{2}{3}}$ and substituting \eqref{V} and \eqref{U} into the Helmholtz equation \eqref{Helm} and boundary condition \eqref{cond}, we immediately obtain the problem for the main term $U_0$ of the outgoing wave:
\begin{subequations}\label{problem1}
	\begin{empheq}[left={\empheqlbrace}]{align}
		&\partial_\nu^2 U_0 + i \partial_\sigma U_0 + \nu \theta(\sigma) U_0 = 0,\\
		&\left. \partial_\nu U_0 \right|_{\nu = 0} = -i\theta(\sigma)\sigma e^{-i\frac{\sigma^3}{3}}.\label{pr1cond}
    \end{empheq}
\end{subequations}

We note that in a remarkable paper by A. Popov \cite{PopovLiPar_en}, which we followed up until now, the exponential on the right-hand side of \eqref{pr1cond} was missing, and thus, his further results require refinement.

\subsection{Formal solution of \eqref{problem1}}
We seek the main term of the attenuation factor $U_0$ in the form of a Fourier-type integral as follows:
\begin{equation}
	U_0(\sigma, \nu) = \frac{1}{2\pi} \int\limits_{-\infty}^{\infty} \widehat{U}_0 (\xi, \nu) e^{i\sigma\xi} d\xi.
\end{equation}
Using the representation for the Heaviside function
\begin{equation}
	\theta(\sigma) = -\frac{i}{2\pi} \int\limits_{-\infty}^{\infty} \frac{e^{i\sigma\xi}}{\xi-i0}  d\xi,
\end{equation}
where $(\xi-i0)^{-1} := \lim\limits_{\varepsilon \to +0} (\xi-i\varepsilon)^{-1}$, and the convolution theorem \cite{GelfandShilov_en}, we arrive at
\begin{subequations}\label{problem2}
	\begin{empheq}[left={\empheqlbrace}]{align}
		&\partial_\nu^2 \widehat{U}_0(\xi, \nu) - \xi \widehat{U}_0(\xi, \nu) + \frac{i\nu}{2\pi} \int\limits_{-\infty}^{\infty} \frac{\widehat{U}_0(t, \nu)}{t - (\xi-i0)} dt = 0, \label{pr2eq}\\
		&\left. \partial_\nu \widehat{U}_0 \right|_{\nu = 0} = I'(\xi).		
	\end{empheq}
\end{subequations}
Here,
\begin{equation}\label{I_def}
	I(\xi) = \int\limits_{0}^{\infty} e^{-i\left(\sigma\xi + \frac{\sigma^3}{3} \right)} d\sigma
\end{equation}
is an \textit{inhomogeneous Airy function} (\ref{Appendix}), and $'$ denotes the differentiation with respect to $\xi$.
The function $I(\xi)$ vanishes in the lower half of the complex plain $\mathbb{C}^-$ as $|\xi|\to \infty$.

Assume that $\widehat{U}_0$ is analytic with respect to $\xi$ and decreases to zero in $\mathbb{C}^-$ as $|\xi|\to \infty$.\footnote{This assumption is a form of causality condition and is equivalent to the condition $U_0(\sigma)=0$ for $\sigma < 0$.}
The integral in \eqref{pr2eq} can be evaluated with the help of Jordan's lemma and residue theorem, and problem \eqref{problem2} takes the form
\begin{subequations}\label{PrF}
	\begin{empheq}[left={\empheqlbrace}]{align}
		&\partial^2_\nu\widehat{U}_0(\xi, \nu) - \left(\xi - \nu \right) \widehat{U}_0(\xi, \nu) = 0, \label{Airyeq} \\
		&\left. \partial_\nu \widehat{U}_0 \right|_{\nu = 0} = I'(\xi).
	\end{empheq}		
\end{subequations}
The desired solution of \eqref{PrF} is
\begin{equation}
	\widehat{U}_0(\xi, \nu) = -\frac{I'(\xi)}{w_1'(\xi)} w_1(\xi - \nu),
\end{equation}
where $w_1$ is the Airy function in Fock's definition (\ref{Appendix}).
We choose this particular solution of the Airy equation \eqref{Airyeq}, guided by the same argument as in Fock \cite{FockDiffProbl_en} and Babich and Kirpichnikova \cite{BabichKirp_en}, as it corresponds to the outgoing wave for the harmonic time-dependence $e^{-ikt}$, which we omit.
Therefore,
\begin{equation}\label{U_0}
	U_0(\sigma, \nu) = -\frac{1}{2\pi} \int\limits_{-\infty}^{\infty} \frac{I'(\xi)}{w_1'(\xi)} w_1(\xi - \nu) e^{i\sigma\xi} d\xi.
\end{equation}
The zeroes of $w_1'$ are located at the upper half-plane; thus, the integrand is analytic with respect to $\xi$ in $\mathbb{C}^-$ and decreases there as $|\xi| \gg 1$ (\ref{Appendix}).
Hence, $U_0 \equiv 0$ as $\sigma \leq 0$, and hereinafter we assume $\sigma > 0$.

The main term of the attenuation factor for the total wavefield is, according to \eqref{sum}, the sum of the main terms of \eqref{U} and \eqref{V}:
\begin{equation}\label{W}
	W_0 = \theta(\sigma) e^{i(\sigma\nu - \frac{\sigma^3}{3})} - \frac{1}{2\pi} \int\limits_{-\infty}^{\infty} \frac{I'(\xi)}{w_1'(\xi)} w_1(\xi - \nu) e^{i\sigma\xi} d\xi.
\end{equation}
This can be equivalently rewritten in the form
\begin{equation}\label{W1}
	W_0 = \frac{1}{2\pi} \int\limits_{-\infty}^{\infty} \left( I(\xi - \nu) - \frac{I'(\xi)}{w_1'(\xi)} w_1(\xi - \nu) \right) e^{i\sigma\xi} d\xi,
\end{equation}
convenient for studying the wavefield in the shadow zone.

\begin{figure}
	\noindent\centering{
	\includegraphics[width=0.7\textwidth]{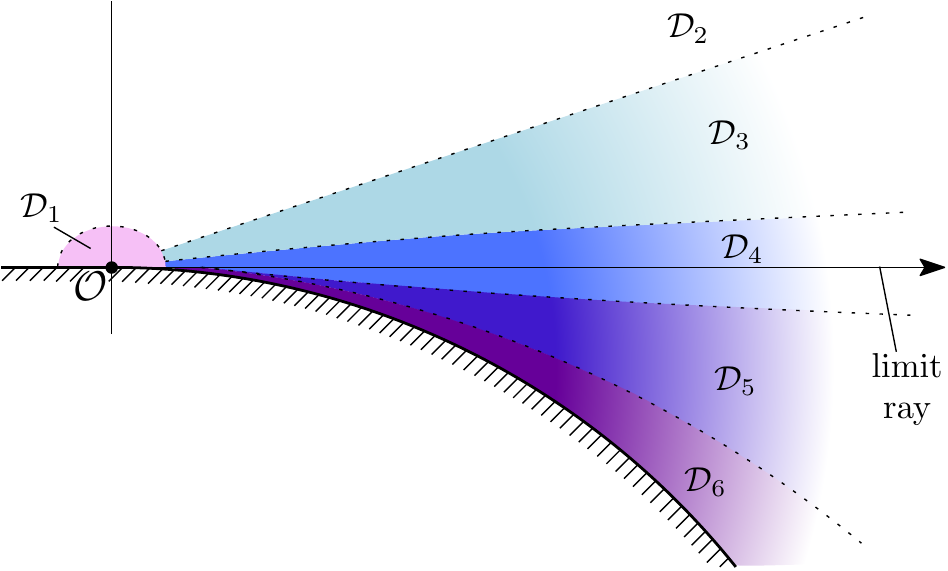}}
	\caption{Schematic sketch of areas under consideration}\label{TotalPic}
\end{figure}

The expression \eqref{U_0} is obtained under the assumption that the inequalities \eqref{upperlim} are satisfied.
When neither $\sigma$ nor $\nu$ is large (the area indicated in Fig. \ref{TotalPic} by $\mathcal{D}_1$), the integral does not allow for a high-frequency asymptotic simplification.
Henceforth, we investigate \eqref{U_0} in areas $\mathcal{D}_{2}$--$\mathcal{D}_6$ where at least one of $\sigma$ and $\nu$ is large,
\begin{equation}\label{guarant}
	\nu + \sigma \gg 1.
\end{equation}
The condition \eqref{guarant} guarantees that all integrals that will be encountered below will contain a large parameter and can be subjected to an asymptotic analysis.

The foregoing analysis relies on a well-developed asymptotic evaluation technique of the integrals of rapidly oscillating functions \cite{Erdelyi_en}.
For different parts of the integration interval, all or some special functions in \eqref{U_0} can be replaced by their approximations.
We will transform expression \eqref{U_0} using the approach of Fock \cite{FockDiffProbl_en}, Brown \cite{Brown} and Babich and Kirpichnikova \cite{BabichKirp_en}, implying the simplification of integrands in small vicinities of critical points of respective phase functions.
We matched the asymptotics of \eqref{U_0} derived in the well-illuminated area $\mathcal{D}_{2}$ with the cylindrical wave \eqref{cylindrical}, which yields an expression for the diffraction coefficient.
The approximation of \eqref{W1}, which we constructed inside the deep shadow area $\mathcal{D}_6$, allows matching with the Friedlander--Keller formulas.

Neighboring areas defined by the corresponding inequalities intersect in the vicinity of their common boundaries, and the respective asymptotic formulas that we construct match in these overlapping zones.

\section{Illuminated area $\mathcal{D}_2$}\label{D2}
Let the observation point be positioned far above the limit ray in the illuminated zone $\mathcal{D}_2$ (Fig.~\ref{TotalPic}), where
\begin{equation}\label{far}
	\nu \gg 1, \quad \nu - \sigma^2 \gg \sigma.
\end{equation}
In polar coordinates, the second inequality means that \eqref{phi}
\begin{equation}\label{farp}
	\varphi \gg (h/k)^{\frac{1}{3}}.
\end{equation}
We separately considered segments of the real axis where the integrand of \eqref{U_0} behaves differently.

\subsection{Segment $-\xi \gg 1$}
On the half-line $-\xi \gg 1$, all special functions in \eqref{U_0} can be replaced by their approximations (\ref{Appendix}).
The integrand takes the form
\begin{multline}\label{-xigg1}
	\frac{I'(\xi)}{w_1'(\xi)} w_1(\xi - \nu) e^{i\sigma\xi} = \left(\frac{\sqrt{\pi} e^{i\Psi_1(\xi) -i\frac{\pi}{4}}}{(\nu - \xi)^\frac{1}{4}} - \frac{e^{i\Psi_2(\xi)}}{(\nu - \xi)^\frac{1}{4} (-\xi)^\frac{9}{4}} \right) \\
	\times \left(1+ O\left((- \xi)^{-\frac{3}{2}}\right) + O\left((\nu - \xi)^{-\frac{3}{2}}\right) \right),
\end{multline}
with the phases
\begin{gather}
		\Psi_1(\xi) = \frac{2}{3} (\nu - \xi)^{\frac{3}{2}} + \sigma\xi, \label{Psi1} \\
		\Psi_2(\xi) = \frac{2}{3} (\nu - \xi)^{\frac{3}{2}} - \frac{2}{3} (- \xi)^{\frac{3}{2}} + \sigma\xi. \label{Psi2}
\end{gather}
The phase $\Psi_1$ has a unique critical point $\xi_1$ where $\Psi_1'(\xi_1)=0$:
\begin{equation}\label{xi_F}
	\xi_1 = \nu - \sigma^2.
\end{equation}
Under the condition \eqref{far}, $\xi_1 \gg 1$, whence the first term on the right-hand side of \eqref{-xigg1} has no critical point on the half-line under consideration and its contribution is negligible.

Now, we address the second term.
The equation for the critical point $\xi_2$ of $\Psi_2$ is
\begin{equation}
	0 = \Psi'_2(\xi_2) \equiv \sigma -(\nu - \xi_2)^{\frac{1}{2}} + (-\xi_2)^{\frac{1}{2}},
\end{equation}
whence $(-\xi)^{\frac{1}{2}} = (\nu - \sigma^2)/(2\sigma)$.
Because the inequality \eqref{far} guarantees that the right-hand side is positive and large, the phase has a critical point
\begin{equation}\label{xi_D}
	\xi_2 = - \left(\frac{\nu - \sigma^2}{2\sigma}\right)^2,
\end{equation}
on the half-line under consideration.

\subsection{Segments $|\xi| < const$}
On the interval $|\xi| < const$, only the function $w_1(\xi - \nu)$ in \eqref{U_0} can be replaced by its approximation \eqref{AsEx_w1}, which gives
\begin{equation}\label{moderate_xi}
	\frac{I'(\xi)}{w_1'(\xi)} w_1(\xi - \nu) e^{i\sigma\xi} = \frac{I'(\xi) e^{i\Psi_1(\xi) + i\frac{\pi}{4}}}{w_1'(\xi) (\nu - \xi)^{\frac{1}{4}}}  \left(1 + O\left((\nu - \xi)^{-\frac{3}{2}}\right)\right),
\end{equation}
with $\Psi_1$ introduced in \eqref{Psi1}.
Under condition \eqref{far} its critical point is far outside the interval and the respective contribution can thus be ignored.

\subsection{Segment $\xi \gg 1$}
We split the half-line $\xi \gg 1$ into three pieces: $1 \ll \xi \ll \nu$, $|\xi-\nu| < const$ and $\xi \gg \nu$.

Where $1 \ll \xi \ll \nu$, we can substitute into \eqref{U_0} asymptotic expressions for $w_1(\xi - \nu)$, $I'(\xi)$ and $w_1'(\xi)$ (\eqref{AsEx_w1} and \eqref{AsEx_I}), and the integrand takes the form
\begin{equation}\label{42}
	\frac{I'(\xi)}{w_1'(\xi)} w_1(\xi - \nu) e^{i\sigma\xi} = -\frac{e^{\Psi_3(\xi) - i\frac{\pi}{4}}}{\xi^{\frac{9}{4}} (\nu-\xi)^{\frac{1}{4}}}  \left(1 + O\left(\xi^{-\frac{3}{2}}\right) + O\left((\nu - \xi)^{-\frac{3}{2}}\right)\right)
\end{equation}
with
\begin{equation}
	\Psi_3(\xi) = i\left(\sigma\xi + \frac{2}{3}(\nu - \xi)^{\frac{3}{2}}\right) - \frac{2}{3}\xi^{\frac{3}{2}}.
\end{equation}

On the interval $|\xi-\nu| < const$, functions $I'(\xi)$ and $w_1'(\xi)$ can be replaced by their approximations (\eqref{AsEx_w1} and \eqref{AsEx_I}), but $w_1(\xi-\nu)$ cannot. Hence, we have
\begin{equation}
	\frac{I'(\xi)}{w_1'(\xi)} w_1(\xi - \nu) e^{i\sigma\xi} = \frac{i}{\xi^{\frac{9}{4}}} w_1(\xi - \nu) e^{\Psi_4(\xi)} \left(1 + O\left(\xi^{-\frac{3}{2}}\right)\right),
\end{equation}
where
\begin{equation}
	\Psi_4(\xi) = i\sigma\xi - \frac{2}{3}\xi^{\frac{3}{2}}.
\end{equation}

Finally, as $\xi \gg \nu$, all special functions can be asymptotically approximated (\eqref{AsEx_w1} and \eqref{AsEx_I}), and the integrand becomes
\begin{equation}
	\frac{I'(\xi)}{w_1'(\xi)} w_1(\xi - \nu) e^{i\sigma\xi} = \frac{ie^{\Psi_5(\xi)}}{\xi^{\frac{9}{4}}(\xi-\nu)^{\frac{1}{4}}} \left(1 + O\left(\xi^{-\frac{3}{2}}\right) + O\left((\xi - \nu)^{-\frac{3}{2}}\right)\right),
\end{equation}
with
\begin{equation}\label{47}
	\Psi_5(\xi) = i\sigma\xi - \frac{2}{3}\left(\xi^{\frac{3}{2}} - (\xi-\nu)^{\frac{3}{2}}\right).
\end{equation}
From formulas \eqref{42}--\eqref{47} we conclude that the contribution of semi-axis $\xi \gg 1$ to \eqref{U_0} is exponentially small.

\subsection{Matching with cylindrical wave}

The analysis above shows that only the critical point \eqref{xi_D}  contributes to \eqref{U_0}.
Then, the standard stationary phase method \cite{Erdelyi_en} yields
\begin{equation}\label{difr}
	U_0 = \frac{2 e^{-i\frac{\pi}{4}}}{\sqrt{\pi \sigma}}\frac{(2\sigma)^4}{(\nu - \sigma^2)^4} \, e^{i\Psi_2(\xi_2)} \left(1 + O\left(\frac{\sigma^3}{(\nu - \sigma^2)^3}\right)\right),
\end{equation}
where
\begin{equation}
	\Psi_2(\xi_2) = \frac{1}{2}\left(\sigma\nu + \frac{\nu^2}{2\sigma} - \frac{\sigma^3}{6}\right).
\end{equation}

Relations \eqref{r} and \eqref{phi} allow us to match the expression \eqref{difr} with a cylindrical wave \eqref{cylindrical} in the area where $\sigma \gg 1$, but the remainder terms in \eqref{r} are small, which can be written as follows:
\begin{equation}
	k(x-s)^2 \ll s.
\end{equation}
The matching easily gives the following formula for the diffraction coefficient $A(\varphi; k)$:
\begin{equation}
	A(\varphi; k) = \sqrt{\frac{2}{\pi}} \frac{h}{k} \frac{2}{\varphi^4} e^{-i\frac{\pi}{4}}.
\end{equation}
This formula is in agreement with the result of A. Popov \cite{PopovLiPar_en}.
Similar to the case of non-tangential incidence (see, e.g., \cite{KiselevRogoff,KamKell,ZloKisWM20}), the diffraction coefficient is linear in the jump of the curvature $h$, but the singularity when approaching to the limit ray is different.

We establish that in the area described by $kr\gg 1$ and \eqref{farp} the total wavefield is the sum of incident and diffracted waves.

\section{Preliminary analysis of Fock-type integrals in the penumbra}\label{PreAn}

Investigations on wavefields in the penumbra for a smooth contour \cite{FockDiffProbl_en,Brown,BabichKirp_en} suggest a transformation of the expression \eqref{U_0}.
First, the classical Airy function $w_1$ (\ref{Appendix}) is connected with inhomogeneous Airy functions $I$ \eqref{I_def} and $H$,
\begin{equation}\label{H_def}
	H(z) = \int\limits_{0}^{\infty} e^{zt - \frac{t^3}{3}} dt,
\end{equation}
by the simple relation
\begin{equation}\label{w_1HI}
	\sqrt{\pi} w_1(z) = i I(z) + H(z).
\end{equation}
This identity allows us to rewrite \eqref{U_0} as follows:
\begin{equation}\label{U_03}
	U_0 = F + D + G.
\end{equation}
Here,
\begin{gather}
	F = \frac{i}{2\sqrt{\pi}} \int\limits_{-\infty}^{0} w_1(\xi - \nu) e^{i\sigma\xi} d\xi, \label{F}\\
	D = - \frac{i}{2\pi} \int\limits_{-\infty}^{0}\frac{H'(\xi)}{w_1'(\xi)} w_1(\xi - \nu) e^{i\sigma\xi} d\xi \label{D}
\end{gather}
and
\begin{equation}\label{G}
	G = - \frac{1}{2\pi} \int\limits_{0}^{\infty} \frac{I'(\xi)}{w_1'(\xi)} w_1(\xi - \nu) e^{i\sigma\xi} d\xi.
\end{equation}
The decomposition \eqref{U_03}, as we will see further, is essential in penumbral areas $\mathcal{D}_4$ and $\mathcal{D}_6$ and will be helpful in the illuminated area $\mathcal{D}_3$ and shadow zone $\mathcal{D}_7$ (Fig. \ref{TotalPic}).

We study functions $F$, $D$ and $G$, which we refer to as Fock-type integrals, separately by applying a specific analytical technique developed by Babich and Kirpichnikova \cite{BabichKirp_en}.
Function $F$ corresponds, by Fock's terminology, to the \textit{Fresnel part} of the wavefield.
Functions $D$ and $G$ are analogous to Fock's \textit{background part} of the wavefield.

\subsection{Function $F$}\label{FSec}
As $\nu \gg 1$, replacing the function $w_1$ by its asymptotic expansion \eqref{AsEx_w1} gives
\begin{equation}\label{F_BK}
	F = -\frac{e^{-i\frac{\pi}{4}}}{2\sqrt{\pi}} \int\limits_{-\infty}^{0} \frac{e^{i\Psi_1(\xi)}}{(\nu-\xi)^{\frac{1}{4}}} \left(1 + O\left((\nu-\xi)^{-\frac{3}{2}}\right) \right)d\xi,
\end{equation}
where the phase $\Psi_1$ introduced in \eqref{Psi1} has the critical point $\xi_1$ \eqref{xi_F}.
When $\xi_1$ is negative and not small, it contributes to the integral; however, when $\xi_1$ is positive and not small, it does not.
These cases correspond to the location of the observation point not too close to the limit ray in the shadow zone and illuminated region, respectively.
Accordingly, these areas are characterized by the inequality
\begin{equation}\label{notFre}
	\nu^{\frac{1}{4}} |\sqrt{\nu} - \sigma| \gg 1,
\end{equation}
which is (\eqref{r}, \eqref{phi}) equivalent to
\begin{equation}
	kr\varphi^2 \gg 1.
\end{equation}
If the critical point of phase $\xi_1$ and the endpoint $0$ merge, then $\xi_1$ contributes to the integral whenever it is negative or positive.
The standard stationary phase method is not applicable; however, the expression \eqref{F_BK} can be rewritten in terms of the Fresnel integral
\begin{equation}\label{FreInt}
	\Phi(z) = \frac{e^{-i\frac{\pi}{4}}}{\sqrt{\pi}} \int\limits_{-\infty}^{z} e^{it^2}dt,
\end{equation}
as follows. Consider a very narrow neighborhood of the limit ray where the inequalities
\begin{equation}\label{penumbra_cond}
	\nu \gg 1, \quad |\sqrt{\nu} - \sigma| \ll 1
\end{equation}
hold true.
In polar coordinates the second one has the form
\begin{equation}
	|\varphi| \ll (h/k)^{\frac{1}{3}}.
\end{equation}
Under the condition \eqref{penumbra_cond}, $|\xi_1| \ll \sqrt{\nu}$, and the main contribution to the integral \eqref{F_BK} is given by the interval $0\le -\xi \ll \sqrt{\nu}$.
Expanding the integrand of \eqref{F_BK} in powers of $1/\nu$ yields
\begin{equation}\label{w1_phase}
	\Psi_1(\xi) = \frac{2}{3} \nu^{\frac{3}{2}} + \xi (\sigma - \sqrt{\nu}) + \frac{\xi^2}{4\sqrt{\nu}} + O\left(\frac{\xi^3}{\nu^\frac{3}{2}} \right).
\end{equation}
On the interval under consideration, the cubic term in $\xi$ and higher-order terms are small.
Taking a quadratic approximation of the phase and main-term approximation of amplitude and extending integration to the half-line $(-\infty-i\varepsilon, 0]$, $\varepsilon > 0$, we derive the following expression for $F$:
\begin{equation}
	F = -\frac{e^{-i\frac{\pi}{4}}}{2\sqrt{\pi}} \frac{e^{i\frac{2}{3}\nu^{\frac{3}{2}}}}{\nu^{\frac{1}{4}}} \int\limits_{-\infty -i\varepsilon}^{0} e^{i\left((\sigma-\sqrt{\nu})\xi + \frac{\xi^2}{4\sqrt{\nu}} \right)} \left(1 + O \left(\frac{\xi^3}{\nu^{\frac{3}{2}}}\right) \right)d\xi.
\end{equation}
This can be immediately rewritten in terms of the Fresnel integral \eqref{FreInt}
\begin{equation}\label{F_penumbra}
	F = - e^{i\Theta} \Phi \left(- \mathrm{Z} \right) \left(1 + O\left((\sqrt{\nu} - \sigma)^3 \right) \right)
\end{equation}
with
\begin{equation}\label{FreArgs}
	\Theta = \frac{2}{3}\nu^\frac{3}{2} - \sqrt{\nu}(\sigma - \sqrt{\nu})^2, \quad
	\mathrm{Z} = \nu^{\frac{1}{4}} (\sqrt{\nu} - \sigma).
\end{equation}
Under the condition \eqref{penumbra_cond}, the remainder terms in \eqref{F_penumbra} are small.

In the area described by \eqref{penumbra_cond}, the phase of the exponential and the argument of the Fresnel integral allow for the following geometrical interpretation (see \eqref{x,y->s,n_str}--\eqref{phi}):
\begin{equation}\label{geomint}
	\Theta \approx k(x-s), \quad \mathrm{Z} \approx \sqrt{\frac{kr}{2}}\,\varphi.
\end{equation}
Variable $\mathrm{Z}$ is typical of wavefields description in areas where waves of different natures merge \cite{BorovikovKinber_en}.
In a part of the area characterized by \eqref{penumbra_cond} (namely, where \eqref{notFre} holds), $\mathrm{Z}$ can be large, and the Fresnel integral allows an asymptotic expansion.

\subsection{Functions $D$ and $G$}\label{DGSec}

We start with the analysis of the function $D$ \eqref{D}.

In the part of the integration interval where $const < \xi \le 0$, the function $w_1(\xi-\nu)$ can be replaced solely by its asymptotic expansion \eqref{AsEx_w1}.
As established in Section \ref{D2}, the derivative of the phase of the integrand vanishes at the point $\xi_1$ \eqref{xi_F}.
Even if the critical point $\xi_1$ lies inside the interval under consideration, it gives no stationary-phase contribution because the integrand does not rapidly oscillate there.

On the half-line $-\xi \gg 1$, all special functions can be replaced by their approximations (\eqref{AsEx_w1} and \eqref{AsEx_H}), which gives
\begin{equation}
	\frac{H'(\xi)}{w_1'(\xi)} w_1(\xi - \nu) e^{i\sigma\xi} = \frac{i e^{i\Psi_2(\xi)}}{(\nu - \xi)^\frac{1}{4} (-\xi)^\frac{9}{4}} \left(1+ O\left((- \xi)^{-\frac{3}{2}}\right) + O\left((\nu - \xi)^{-\frac{3}{2}}\right) \right).
\end{equation}
Here, $\Psi_2$ is introduced in \eqref{Psi2}.
In Section \ref{D2}, we show that under the condition \eqref{far}, phase $\Psi_2$ has one critical point $\xi_2$ \eqref{xi_D}, of which the contribution to $U_0$ has the form \eqref{difr}.

Now, we consider the vicinity of the limit ray, where the inequalities
\begin{equation}\label{close}
	\nu \gg 1 , \quad \left|\sqrt{\nu} - \sigma \right| \ll \nu^{\frac{1}{8}}
\end{equation}
hold true.
In polar coordinates, they read
\begin{equation}
	kh^2r^3\gg 1, \quad kr\varphi^4 \ll (hr)^2
\end{equation}
The area characterized by \eqref{close} is wider than that described by \eqref{penumbra_cond}.

The inequality \eqref{close} implies that $-\xi_2 \ll \nu^{\frac{1}{4}}$, whence the main contribution to $D$ is given by the segment $0 \le -\xi \ll \nu^{\frac{1}{4}}$.
We replace $w_1(\xi-\nu)$ by its approximation \eqref{AsEx_w1}:
\begin{equation}
	\frac{H'(\xi)}{w_1'(\xi)} w_1(\xi - \nu) e^{i\sigma\xi} =\frac{H'(\xi)}{w_1'(\xi)} \frac{e^{i\Psi_1(\xi)}}{(\nu - \xi)^{\frac{1}{4}}}  \left(1+O\left((\nu - \xi)^{-\frac{3}{2}}\right)\right).
\end{equation}
Here, $\Psi_1$ is introduced in \eqref{Psi1}. Then, we expand the integrand in powers of $1/\nu$.
Phase $\Psi_1$ is rewritten as \eqref{w1_phase}.
As $|\xi| \ll \nu^{\frac{1}{4}}$, the quadratic term in $\xi$ and higher-order terms in \eqref{w1_phase} are small and can be discarded.
Therefore, we rewrite the integrand as follows:
\begin{equation}\label{Dmodint}
	\frac{e^{i\frac{2}{3}\nu^{\frac{3}{2}}}}{\nu^{\frac{1}{4}}} \frac{H'(\xi)}{w_1'(\xi)} e^{i(\sigma - \sqrt{\nu})\xi}
\left(1 + O\left(\frac{\xi^2}{\sqrt{\nu}}\right) + O\left(\frac{\xi}{\nu}\right)+O\left((\nu-\xi)^{-\frac{3}{2}}\right)\right).
\end{equation}
On the half-line $-\xi \gg 1$, the functions $w_1$ and $H$ can be replaced by their approximations \eqref{AsEx_w1} and \eqref{AsEx_H}, respectively, and we arrive at an integrand with the phase $\widetilde{\Psi}_2 = -\frac{2}{3}(-\xi)^\frac{3}{2} + (\sigma - \sqrt{\nu})\xi$.
A straightforward calculation (similar to that concerning the phase $\Psi_2$; see \eqref{Psi2}--\eqref{xi_D}) shows that $\widetilde{\Psi}_2$ has exactly one critical point
\begin{equation}\label{xi_*}
	\widetilde{\xi}_2=-(\sqrt{\nu} - \sigma)^2,
\end{equation}
provided that $\sqrt{\nu} - \sigma \gg 1$, cf. \eqref{far}.
The condition \eqref{close} implies that $-\widetilde{\xi}_2 \ll \nu^{\frac{1}{4}}$, whence the integral of \eqref{Dmodint} over the half-line $-\xi \ge const\,\nu^{\frac{1}{4}}$ is negligible.
Therefore, under the conditions \eqref{close}, function $D$ has the following asymptotic representation:
\begin{equation}\label{Dclose}
	D = \frac{e^{-i\frac{\pi}{4}}}{2\pi} \frac{e^{i\frac{2}{3} \nu^{\frac{3}{2}}}}{\nu^{\frac{1}{4}}} \int\limits_{-\infty}^{0} \frac{H'(\xi)}{w_1'(\xi)} e^{i\xi(\sigma - \sqrt{\nu})} d\xi \left(1+o\left(1\right)\right).
\end{equation}

A similar manipulation with $G$ yields
\begin{equation}\label{Gclose}
	G = -\frac{e^{i\frac{\pi}{4}}}{2\pi} \frac{e^{i\frac{2}{3} \nu^{\frac{3}{2}}}}{\nu^{\frac{1}{4}}} \int\limits_{0}^{\infty} \frac{I'(\xi)}{w_1'(\xi)} e^{i\xi(\sigma - \sqrt{\nu})} d\xi \left(1+o(1)\right).
\end{equation}

\section{Penumbral area $\mathcal{D}_4$}

Let the observation point be positioned inside the penumbral region $\mathcal{D}_4$ characterized by inequalities \eqref{penumbra_cond}.
The results in Section \ref{PreAn} imply the following representation of \eqref{W}:
\begin{multline}\label{S3}
	W_0 = e^{i\Theta} \Phi \left(\mathrm{Z} \right)  \left(1 + O\left((\sqrt{\nu} - \sigma)^3 \right) \right) \\
	+\frac{e^{-i\frac{\pi}{4}}}{2\pi} \frac{e^{i\frac{2}{3} \nu^{\frac{3}{2}}}}{\nu^{\frac{1}{4}}}
	\left(\int\limits_{-\infty}^{0} \frac{H'(\xi)}{w_1'(\xi)} e^{i\xi(\sigma - \sqrt{\nu})} d\xi - i \int\limits_{0}^{\infty} \frac{I'(\xi)}{w_1'(\xi)} e^{i\xi(\sigma - \sqrt{\nu})} d\xi \right)\left(1+o(1)\right),
\end{multline}
where $\Theta$ and $\mathrm{Z}$ were introduced in \eqref{FreArgs}.

The first term on the right-hand side of \eqref{S3} is the Fresnel part of the wavefield, which is exactly the same as that in the Fock case \cite{FockDiffProbl_en,Brown,BabichKirp_en}, and, like theirs, does not depend on the curvature of the contour \eqref{geomint}.
The second term in \eqref{S3} is analogous to Fock's background part of the wavefield and transforms into it after replacing functions $I$ and $H$ with the classical Airy functions $v$ and $w_2$ (\ref{Appendix}), respectively.
Similarly to Fock's case, the Fresnel part is superimposed on the background.

The inequality \eqref{penumbra_cond} characterizes the transition zone where $u^\text{inc}$ cannot be considered as an individual wave.

\section{Illuminated area $\mathcal{D}_3$}\label{D3}

We resume consideration of the illuminated area, addressing its part $\mathcal{D}_3$ (Fig. \ref{TotalPic}), where conditions \eqref{notFre} and \eqref{close} hold and
\begin{equation}\label{phi>0}
	\varphi > 0.
\end{equation}
Here, we employ the representation \eqref{U_03} for the outgoing wavefield.
Particularly, \eqref{close} describes the area wider than that characterized by \eqref{penumbra_cond}.
Thus, the representation \eqref{F_penumbra} for the function $F$ \eqref{F} is not applicable in the entire area $\mathcal{D}_3$.

We consider function $F$ \eqref{F}, starting with its approximation \eqref{F_BK}.
Although the phase $\Psi_1$ \eqref{Psi1} of the integrand in \eqref{F_BK} has a critical point $\xi_1$ \eqref{xi_F}, it is positive and large when \eqref{phi>0} and \eqref{notFre} hold and does not contribute to $F$.
Thus, it is sufficient to consider only a small neighborhood of endpoint $0$.
As in Section \ref{FSec}, we expand the integrand of \eqref{F_BK} in powers of $1/\nu$, and the phase takes the form \eqref{w1_phase}.
Now, because $|\xi|$ is small, we discard not only the cubic term in $\xi$ and higher-order terms but also the quadratic term.
We retain the linear term in the phase and then extend the integration to the half-line $(-\infty-i\varepsilon, 0]$, $\varepsilon > 0$.
This manipulations result in the following approximation for $F$:
\begin{equation}\label{F_Q_ill}
	F = \frac{e^{-i\frac{\pi}{4}}}{2\sqrt{\pi}} \frac{e^{i\frac{2}{3}\nu^{\frac{3}{2}}}}{\nu^{\frac{1}{4}}} \int\limits_{-\infty-i\varepsilon}^{0} e^{i(\sigma-\sqrt{\nu})\xi} d\xi \left(1 + o(1) \right).
\end{equation}
Although the integral \eqref{F_Q_ill} can be explicitly evaluated, it is convenient to leave it as it is.

We now sum the functions $F$, $D$ and $G$, utilizing their representations \eqref{F_Q_ill}, \eqref{Dclose} and \eqref{Gclose}, and obtain the expression for $U_0$.
We deform the integration contours in \eqref{F_Q_ill} and \eqref{Gclose} to the half-line $\arg \xi = -i\pi/3$.
Accounting \eqref{w_1HI} allows
\begin{multline}
	F + G = \frac{e^{-i\frac{\pi}{4}}}{2\pi} \frac{e^{i\frac{2}{3}\nu^{\frac{3}{2}}}}{\nu^{\frac{1}{4}}} \int\limits_{0}^{\infty e^{-i\pi/3}} \left(\sqrt{\pi} - i \frac{I'(\xi)}{w_1'(\xi)} \right)e^{i(\sigma-\sqrt{\nu})\xi} d\xi \left(1 + o(1) \right) \\
	=  \frac{e^{-i\frac{\pi}{4}}}{2\pi} \frac{e^{i\frac{2}{3}\nu^{\frac{3}{2}}}}{\nu^{\frac{1}{4}}} \int\limits_{0}^{\infty e^{-i\pi/3}} \frac{H'(\xi)}{w_1'(\xi)} e^{i(\sigma-\sqrt{\nu})\xi} d\xi \left(1 + o(1) \right).
\end{multline}
Finally, using the expression \eqref{Dclose} for $D$, we come up with
\begin{equation}\label{U_D3}
	U_0 = F+D+G= \frac{e^{-i\frac{\pi}{4}}}{2\pi} \frac{e^{i\frac{2}{3} \nu^{\frac{3}{2}}}}{\nu^{\frac{1}{4}}} \left(\int\limits_{-\infty}^{0} + \int\limits_{0}^{\infty e^{-i\pi/3}}\right) \frac{H'(\xi)}{w_1'(\xi)} e^{i\xi(\sigma - \sqrt{\nu})} d\xi \left(1+o\left(1\right)\right).
\end{equation}

We consider the expression \eqref{U_D3} well apart from the limit ray in the illuminated area, where the condition \eqref{far} is satisfied.

On the segments of integration contours where $|\xi| \gg 1$, the special functions on the right-hand side of \eqref{U_D3} can be replaced by their approximations (\ref{Appendix}).
Now, for the integral over $[0, \infty e^{-i\pi/3})$, it can be easily shown that its asymptotics is given by the contribution of the endpoint $0$.
We consider the integral over $(-\infty, 0]$.
As follows from the results in Section \ref{DGSec}, its asymptotics is given by the contributions of the critical point \eqref{xi_*} and the endpoint $0$, and the latter fully cancels with the aforementioned approximation of the integral over $[0, \infty e^{-i\pi/3})$.
Consequently, the asymptotics of \eqref{U_D3} is given by the critical point of the phase:
\begin{equation}
	U_0 = \frac{2 e^{-i\frac{\pi}{4}}}{\sqrt{\pi} \nu^{\frac{1}{4}} (\sqrt{\nu} - \sigma)^4} e^{i\left(\frac{2}{3} \nu^{\frac{3}{2}} + \frac{(\sqrt{\nu} - \sigma)^3}{3} \right)}(1+o(1)).
\end{equation}
When the conditions \eqref{far} and \eqref{close} are satisfied, this formula agrees with the expression \eqref{difr}, which matches the diffracted wave.

\section{Shadow area $\mathcal{D}_5$}\label{D7}
Now, we consider the wavefield in the shadow area $\mathcal{D}_5$,
\begin{equation}\label{phi<0}
    \varphi < 0,
\end{equation}
described by the inequalities \eqref{notFre} and \eqref{close} (Fig. \ref{TotalPic}).

\subsection{Preliminary transformation of the integral \eqref{W1}}
Using the relation \eqref{w_1HI} connecting functions $w_1$, $I$ and $H$, we rewrite the expression \eqref{W1} for the attenuation factor of the total wavefield as follows:
\begin{multline}\label{W0D7}
	W_0 = \left(\frac{i}{2\pi} \int\limits_{-\infty}^{0} H(\xi - \nu) e^{i\sigma\xi} d\xi + \frac{1}{2\pi} \int\limits_{0}^{\infty} I(\xi - \nu) e^{i\sigma\xi} d\xi \right) \\
	- \frac{1}{2\pi} \int\limits_{0}^{\infty} \frac{I'(\xi)}{w_1'(\xi)} w_1(\xi - \nu) e^{i\sigma\xi} d\xi - \frac{i}{2\pi} \int\limits_{-\infty}^{0} \frac{H'(\xi)}{w_1'(\xi)} w_1(\xi - \nu) e^{i\sigma\xi} d\xi.
\end{multline}

We observe that
\begin{equation}\label{w1_pi3}
	\frac{i}{2\pi}\int\limits_{-\infty}^{0} H(\xi - \nu) e^{i\sigma\xi} d\xi + \frac{1}{2\pi}\int\limits_{0}^{\infty} I(\xi - \nu) e^{i\sigma\xi} d\xi  = -\frac{i}{2\sqrt{\pi}} \int\limits_{0}^{\infty e^{i\pi/3}} w_1(\xi - \nu) e^{i\sigma\xi} d\xi.
\end{equation}

\subsection{Investigation of the right-hand side of \eqref{w1_pi3}}

Let us transform the integral on the right-hand side of \eqref{w1_pi3}.
Using the asymptotic formula \eqref{AsEx_w1} for $w_1(\xi - \nu)$, we observe that the integrand decays exponentially on segments $|\xi - \nu| < const$ and $|\xi|\gg \nu$ as $|\xi|$ increases.
Hence, up to an exponentially small error, it is sufficient to deal with the segment $0\le |\xi| \ll \nu$, whence
\begin{equation}\label{w1_pre}
	-\frac{i}{2\sqrt{\pi}} \int\limits_{0}^{\infty e^{i\frac{\pi}{3}}} w_1(\xi - \nu) e^{i\sigma\xi} d\xi = \frac{e^{-i\frac{\pi}{4}}}{2\sqrt{\pi}} \int\limits_{\substack{0\le |\xi| \ll \nu\\ \arg\xi=\pi/3}}\frac{e^{i\Psi_1(\xi)}}{(\nu-\xi)^{\frac{1}{4}}} \left(1 + O\left((\nu-\xi)^{-\frac{3}{2}}\right) \right) d\xi
\end{equation}
with $\Psi_1$ introduced in \eqref{Psi1}.

Next, we expanded the integrand in \eqref{w1_pre} in powers of $1/\nu$, where $\Psi_1$ takes the form of \eqref{w1_phase}.
We observed an $\Im\Psi_1 > 0$ on the integration interval.
Indeed, in accordance with \eqref{phi<0}, $\Im \xi (\sigma-\sqrt{\nu}) > 0$.
Evidently, the imaginary part of the quadratic term in $\xi$ is also positive.
Thus, the integrand exponentially decays as $|\xi|$ increases, and the asymptotics of the integral is given by the contribution of the endpoint $0$.
We retain only the term linear in $\xi$ in the phase and finally obtain
\begin{equation}\label{w1_Q}
	w_1(\xi-\nu)e^{i\sigma\xi} = \frac{e^{i\frac{2}{3}\nu^{\frac{3}{2}}+i\frac{\pi}{4}}}{\nu^{\frac{1}{4}}} e^{i(\sigma-\sqrt{\nu})\xi} \left(1 + Q\left(\frac{\xi}{\nu^{\frac{1}{4}}}, \frac{1}{\nu^{\frac{1}{4}}} \right)\right).
\end{equation}
Here, $Q$ is the expansion in the positive integer powers of $\xi/\nu^\frac{1}{4}$ and the non-negative integer powers of $1/\nu^{\frac{1}{4}}$:
\begin{equation}\label{Q}
	Q = \frac{\xi}{4{\nu}^{\frac{1}{4}}} \left(\frac{1}{\nu^{\frac{1}{4}}}\right)^{3} + \frac{1}{4} \left(\frac{\xi}{{\nu}^{\frac{1}{4}}}\right)^2 + \frac{5}{32} \left(\frac{\xi}{{\nu}^{\frac{1}{4}}}\right)^2 \left(\frac{1}{\nu^{\frac{1}{4}}}\right)^6 + \ldots,
\end{equation}
Expressions similar to \eqref{w1_Q}--\eqref{Q} can be found in \cite{BabichKirp_en}.
Substituting the expansion \eqref{w1_Q} into \eqref{w1_pre} and then extending the integration to the half-line $\arg\xi = \pi/3$, we obtain, with superpower accuracy,
\begin{equation}\label{w1_Q_sh}
	-\frac{i}{2\sqrt{\pi}} \int\limits_{0}^{\infty e^{i\pi/3}} w_1(\xi - \nu) e^{i\sigma\xi} d\xi = \frac{e^{-i\frac{\pi}{4}}}{2\sqrt{\pi}} \frac{e^{i\frac{2}{3}\nu^{\frac{3}{2}}}}{\nu^{\frac{1}{4}}}  \int\limits_{0}^{\infty e^{i\pi/3}} e^{i(\sigma-\sqrt{\nu})\xi} \, \left(1 + Q\right) d\xi.
\end{equation}

\subsection{Asymptotics of \eqref{W0D7}}
Similarly, with superpower accuracy, we derive the approximation for the second term on the right-hand side of \eqref{W0D7}:
\begin{equation}\label{G_Q_sh}
	-\frac{1}{2\pi}\int\limits_{0}^{\infty}\frac{I'(\xi)}{w_1'(\xi)} w_1(\xi - \nu) e^{i\sigma\xi} d\xi = -\frac{e^{i\pi/4}}{2\pi} \frac{e^{i\frac{2}{3}\nu^{\frac{3}{2}}}}{\nu^{\frac{1}{4}}} \int\limits_{0}^{\infty e^{i\pi/4}}\frac{I'(\xi)}{w_1'(\xi)} e^{i(\sigma-\sqrt{\nu})\xi} \, \left(1 + Q\right) d\xi,
\end{equation}
and for the third term:
\begin{equation}\label{D_Q_sh}
    - \frac{i}{2\pi} \int\limits_{-\infty}^{0} \frac{H'(\xi)}{w_1'(\xi)} w_1(\xi - \nu) e^{i\sigma\xi} d\xi = \frac{e^{-i\frac{\pi}{4}}}{2\pi} \frac{e^{i\frac{2}{3} \nu^{\frac{3}{2}}}}{\nu^{\frac{1}{4}}} \int\limits_{-\infty}^{0} \frac{H'(\xi)}{w_1'(\xi)} e^{i\xi(\sigma - \sqrt{\nu})} \, \left(1 + Q\right) d\xi.
\end{equation}

We now return to \eqref{W0D7}.
Summing the right-hand sides of \eqref{w1_Q_sh} and \eqref{G_Q_sh} with the help of \eqref{w_1HI}, we derive
\begin{multline}\label{I+1}
	\frac{e^{-i\frac{\pi}{4}}}{2\pi} \frac{e^{i\frac{2}{3}\nu^{\frac{3}{2}}}}{\nu^{\frac{1}{4}}} \int\limits_{0}^{\infty e^{i\pi/4}} \left(\sqrt{\pi} - i\frac{I'(\xi)}{w_1'(\xi)} \right) e^{i(\sigma-\sqrt{\nu})\xi} \left(1 + Q \right) d\xi \\
	= \frac{e^{-i\frac{\pi}{4}}}{2\pi} \frac{e^{i\frac{2}{3}\nu^{\frac{3}{2}}}}{\nu^{\frac{1}{4}}} \int\limits_{0}^{\infty e^{i\pi/4}} \frac{H'(\xi)}{w_1'(\xi)} e^{i(\sigma-\sqrt{\nu})\xi} \, \left(1 + Q\right) d\xi.
\end{multline}
Employing equation \eqref{D_Q_sh}, we finally come up with the formula
\begin{equation}\label{W0D7f}
	W_0 = \frac{e^{-i\frac{\pi}{4}}}{2\pi} \frac{e^{i\frac{2}{3}\nu^{\frac{3}{2}}}}{\nu^{\frac{1}{4}}} \left(\int\limits_{-\infty}^{0} + \int\limits_{0}^{\infty e^{i\pi/4}}\right) \frac{H'(\xi)}{w_1'(\xi)} e^{i(\sigma - \sqrt{\nu})\xi} \, \left(1 + Q\right) d\xi
\end{equation}
describing $W_0$ with superpower accuracy.

The integrand in \eqref{W0D7f} has an infinite number of poles at zeroes $\{\zeta_j\}_{j=1}^{\infty}$ of function  $w_1'$ (\ref{Appendix}) located at $\mathbb{C}^+$.
When the observation point goes from the shadowed portion of penumbra deeper into the shadow, and $-\varphi \gg (h/k)^{\frac{1}{3}}$ (i.e. $\sigma - \sqrt{\nu} \gg 1$; see \eqref{phi}), the evaluation of the integral by residues becomes feasible.
We delayed the discussion of such a representation to Section \ref{D8}.

\section{Deep shadow area $\mathcal{D}_6$}\label{D8}
We address the wavefield in the area $\mathcal{D}_6$ in the vicinity of contour $\mathcal{C}$ (Fig. \ref{TotalPic}), as described by
\begin{equation}
	\sigma \gg 1, \quad \sigma^2 - \nu \gg \sigma.
\end{equation}
In polar coordinates, this reads
\begin{equation}
	kh^2 r^3 \gg 1, \quad -\varphi \gg (h/k)^{\frac{1}{3}},
\end{equation}
cf. \eqref{far}.
Further calculation is much in the same way as that presented by Babich and Kirpichnikova \cite{BabichKirp_en}.

The integrand in \eqref{W1} has an infinite number of poles at zeroes $\{\zeta_j\}_{j=1}^{\infty}$ of $w_1'(\xi)$, which are located at the half-line $\arg \xi = \pi/3$ (\ref{Appendix}).
The asymptotic formulas for functions $I$ and $w_1$ and their derivatives (\eqref{AsEx_w1} and \eqref{AsEx_I}) shows that integrand decays as $|\xi|^{-1}$ when $|\Re \xi| \to \infty$ and $\Im \xi > 0$.
Raising the integration contour allows for the representation
\begin{equation}\label{W_res+}
	W_0 = \sum\limits_{j=1}^{N}A_j w_1(\zeta_j - \nu)e^{i\sigma\zeta_j} + \mathscr{E},
\end{equation}
with
\begin{equation}
	\mathscr{E} = \frac{e^{-\varepsilon \sigma}}{2\pi} \int\limits_{-\infty}^{\infty} \left(I(\xi+ i\varepsilon - \nu) - \frac{I'(\xi+ i\varepsilon)}{w_1'(\xi+ i\varepsilon)} w_1(\xi+ i\varepsilon - \nu) \right) e^{i\sigma \xi} d\xi,
\end{equation}
where $\Im \zeta_N < \varepsilon < \Im \zeta_{N+1}$, $N$ is a positive integer, and
\begin{equation}\label{W_res++}
	A_j= - \frac{iI'(\zeta_j)}{\zeta_j w_1(\zeta_j)}.
\end{equation}
Similarly  to the Fock case, the remainder $\mathscr{E}$ is of a smaller order than each of the residues: $\mathscr{E}$ decays as  $e^{-\varepsilon \sigma}$, whereas the residues decay as $e^{- \sigma \Im\zeta_j}$, and $0<\Im\zeta_j < \varepsilon$ \cite{BabichKirp_en}.
Constants $A_j$, $j = 1, \ldots, N$, can be interpreted as the excitation coefficients of creeping modes, and
they differ from those found for the Fock problem by Babich and Kirpichnikova \cite{BabichKirp_en}.

At a distance from the contour $\mathcal{C}$, such that $\nu \gg 1$, functions $w_1(\zeta_j - \nu)$ can be replaced by their approximations \eqref{AsEx_w1}, which yields
\begin{equation}\label{W0as}
	W_0 = \frac{e^{i\frac{\pi}{4}}}{\nu^{\frac{1}{4}}} \sum\limits_{j=1}^{N} A_j e^{i\left(\frac{2}{3}\nu^{\frac{3}{2}} + \zeta_j (\sigma -\sqrt{\nu})\right)} \left(1 + O\left(\frac{1}{\sqrt{\nu}}\right)\right) + \mathscr{E}.
\end{equation}
Accounting the equations \eqref{w_1HI} and $w_1'(\zeta_1)=0$, one can easily observe that $I'(\zeta_j) = iH'(\zeta_j)$ and rewrite formula \eqref{W_res++} as
\begin{equation}\label{W_res++0}
	A_j= - \frac{iH'(\zeta_j)}{\zeta_j w_1(\zeta_j)},
\end{equation}
which makes it possible to indicate that \eqref{W0as} agrees with the representation for $W_0$ resulting from the evaluation of the integral in \eqref{W0D7} by residues.

Moreover, the expression \eqref{W0as} matches with the Friedlander--Keller asymptotic formulas (see, e.g., \cite{BabichKirp_en}).

\section{Conclusions}

We have explored the simplest problem of diffraction on non-smooth contours, where the incidence is tangential and the contour curvature changes with jump.
We hope that the developed technique will be useful in the study of similar problems with concave--convex transitions, which attract significant attention in the smooth case (see, e.g., \cite{HewOckSmyshlWM19,Kazakov03,Nakamura89,OckTew12,OckTew21,PopovMM79_en,PopovMM82_en,PopovMM86_en,Smyshl91_en,SmyshlKamAA22_en}).

Our consideration, similarly to the pioneering research by Fock \cite{Fock0,FockDiffProbl_en}, was limited to the small area characterized by inequalities \eqref{upperlim} resulted from the analysis of approximations which led to the parabolic equation.
In subsequent papers \cite{Brown,BabichKirp_en,Busl,Buldyrev_Lyalinov,Melrose86,Tew+00,Hewett}, steps were taken towards describing the field at a larger distance. Perhaps, these techniques will be helpful in the further study of the Malyuzhinets---Popov problem.

\section*{Acknowledgements}
This work was supported by the Russian Science Foundation grant 22-21-00557.

The authors are indebted to Vladimir E. Petrov and Alexey V. Popov for their helpful discussion and to the anonymous reviewer for a number of useful remarks.

\appendix
\section{Airy and inhomogeneous Airy functions \label{Appendix}}
Here, some properties of Airy and inhomogeneous Airy functions \cite{BabichBuldyrev_en,NIST} are briefly summarized.

\begin{figure}
	\noindent\centering{
	\includegraphics[width=0.9\textwidth]{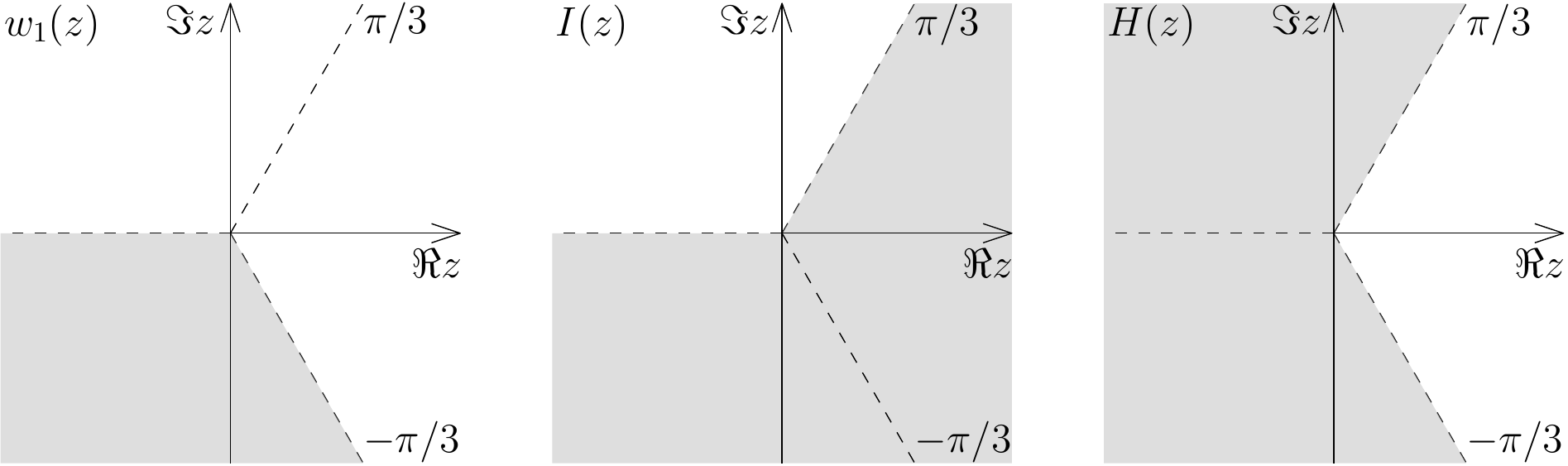}}
	\caption{Behavior of functions $w_1(z)$, $I(z)$ and $H(z)$ in the complex $z$-plane}\label{ChA}
\end{figure}

The notation $w_1(z)$ was introduced by Fock \cite{FockDiffProbl_en} for a particular solution to the homogeneous Airy equation:
\begin{equation}
	w''(z) - z w(z) = 0,
\end{equation}
defined by the integral representation
\begin{equation}\label{w1_def}
	w_1(z) = \frac{1}{\sqrt{\pi}} \int\limits_{\gamma} e^{zt-\frac{t^3}{3}}dt,
\end{equation}
where contour $\gamma$ goes from infinity to zero along the line $\arg t = -2\pi/3$ and then from zero to infinity along the line $\arg t = 0$.
Function $w_1$ is related to another common notation \cite{NIST} as follows:
\begin{equation}
	w_1(z) = 2 \sqrt{\pi} e^{i\frac{\pi}{6}} \mathrm{Ai}\left(z e^{i\frac{2\pi}{3}} \right).
\end{equation}
Both function $w_1$ and its derivative $w_1'$ have an infinite number of simple zeroes located on the line $\arg z = \pi/3$.
The asymptotic expansion of $w_1$ is \cite{BabichBuldyrev_en}
\begin{equation}\label{AsEx_w1}
	\begin{gathered}
		w_{1}(z) = z^{-\frac{1}{4}} e^{\frac{2}{3} z^{\frac{3}{2}}}\left(1+O\left(z^{-\frac{3}{2}}\right)\right), \quad -\frac{4\pi}{3} \le \arg z \le 0,\\
		w_{1}(z) = \left(z^{-\frac{1}{4}} e^{\frac{2}{3} z^{\frac{3}{2}}} + i z^{-\frac{1}{4}} e^{-\frac{2}{3} z^{\frac{3}{2}}} \right)\left(1+O\left(z^{-\frac{3}{2}}\right)\right), \quad 0 < \arg z < \frac{2\pi}{3}.
	\end{gathered}
\end{equation}
In Fig. \ref{ChA}, the sectors where $w_1$ decreases when $|z|\gg 1$ are shown in grey, those where it increases in white and the lines where it oscillates are dashed.

Classical Russian research \cite{FockDiffProbl_en,BabichKirp_en,BabichBuldyrev_en} has also used the following notations:
\begin{equation}
	w_2(z) = \overline{w_1(\overline{z})}, \quad v(z) = \frac{1}{2i} (w_1(z)+w_2(z)),
\end{equation}	
where $\bar{\,\,\,}\!\bar{\,\,\,}$ stands for a complex conjugation.

Functions $I(z)$ \eqref{I_def} and $H(z)$ \eqref{H_def} are inhomogeneous Airy functions.
They can be expressed through the standard inhomogeneous Airy function (also called the Scorer function; see \cite{NIST}):
\begin{equation}
	\mathrm{Hi}(z) = \frac{1}{\pi}\int\limits_0^\infty e^{zt - \frac{t^3}{3}} dt
\end{equation}
solving the inhomogeneous Airy equation
\begin{equation}
	w''(z) - z w(z) = \frac{1}{\pi}
\end{equation}
as follows:
\begin{gather}
	I(z) = \pi e^{-i\frac{\pi}{6}} \mathrm{Hi}\left(z e^{-i\frac{2\pi}{3}}\right), \quad H(z) = \pi\mathrm{Hi}(z).
\end{gather}

The asymptotic approximations of $H$ and $I$ are as follows:
\begin{equation}\label{AsEx_H}
	\begin{gathered}
		H(z) = -\frac{1}{z} \left(1 + O\left(z^{-2}\right)\right),
		\quad -\frac{4\pi}{3} < \arg z < -\frac{2\pi}{3},\\
		H(z) = -\frac{1}{z} \left(1 + O\left(z^{-2}\right)\right)
		+\sqrt{\pi}\,z^{-\frac{1}{4}} e^{\frac{2}{3}z^{\frac{3}{2}}} \left(1 + O\left(z^{-\frac{3}{2}}\right)\right),
		\quad -\frac{2\pi}{3} \leq \arg z \leq \frac{2\pi}{3};
	\end{gathered}
\end{equation}
\begin{equation}\label{AsEx_I}
	\begin{gathered}
		I(z) = -\frac{i}{z} \left(1 + O\left(z^{-2}\right)\right), \quad -\frac{2\pi}{3} < \arg z < 0,\\
		I(z) = -\frac{i}{z} \left(1 + O\left(z^{-2}\right)\right) + \sqrt{\pi}\, z^{-\frac{1}{4}}e^{-\frac{2}{3}z^{\frac{3}{2}}} \left(1 + O\left(z^{-\frac{3}{2}}\right)\right), \quad 0 \leq \arg z \leq \frac{4\pi}{3}.
	\end{gathered}
\end{equation}
In Fig. \ref{ChA}, the sectors where the function under consideration decreases when $|z|\gg 1$ are shown in gray, those where it increases in white and the lines where it oscillates are dashed.

The asymptotic formulas \eqref{AsEx_w1}, \eqref{AsEx_H} and \eqref{AsEx_I} allow differentiation.

The aforementioned functions are all analytic in the complex plane.


\end{document}